\documentclass[12pt]{article}
\usepackage{amsfonts}

\newcommand{\CURL}{\tt{CURL}}
\newcommand{\DIV}{\tt{DIV}}
\newcommand{\GRAD}{\tt{GRAD}}
\newcommand{\sCURL}{\sf{\bf CURL}}

\newcommand{\curl}{\tt{curl}}
\newcommand{\tdiv}{\tt{div}}

\begin{document}

\begin{titlepage}
\vskip 2cm
\begin{center}
{\Large\bf On a group-theoretical approach to the curl operator}\footnote{
 {\tt jramos@phys.ualberta.ca,}{\tt montigny@phys.ualberta.ca,}{\tt khanna@phys.ualberta.ca} }
\vskip 3cm
{\bf
J. Ramos$^{a}$, M. de Montigny$^{a,b}$, F.C. Khanna$^{a,c}$ \\}
\vskip 5pt
{\sl $^a$Theoretical Physics Institute, University of Alberta,\\
Edmonton, Alberta, Canada T6G 2J1\\}
\vskip 2pt
{\sl $^b$Campus Saint-Jean, University of Alberta, \\
Edmonton, Alberta, Canada T6C 4G9\\}
\vskip 2pt
{\sl $^c$TRIUMF, 4004, Westbrook Mall,\\
Vancouver, British Columbia, Canada  V6T 2A3\\}
\vskip 2pt
\end{center}
\vskip .5cm
\rm
\begin{abstract}
We utilize group-theoretical methods to develop a matrix representation of differential operators that act on tensors of any rank. 
 In particular, we concentrate on the matrix formulation of the curl operator. A self-adjoint matrix of the 
curl operator is constructed and its action is extended to a complex plane. This scheme allows us to obtain properties, similar to those 
 of the traditional curl operator. 
\end{abstract}

\end{titlepage}

\section{Introduction}
Many areas of physics, like electromagnetism, hydrodynamics and plasma physics, make use of
 differential operators such as the divergence, the gradient and the curl.  Matrix representations of these differential operators have been
  developed by several authors \cite{Chen, Vis, Chang}.
Efficient techniques based on this matrix formulation have been used to simplify computations in applied physics, vector analysis
and manipulation of the differential operators in boundary value problems \cite{Chen}. The matrix technique
provides an elegant representation of Maxwell and vector wave equations.  Within general orthogonal systems, these operators are represented by matrices that act on vectors and 
whose dimensions do not exceed 3 $\times$ 3. With these matrices, complicated vector operations can be presented in a simple form 
\cite{Vis, Chang}. 

In the present paper, we provide a matrix formalism for the differential operators which is based on group theory and the quantum theory of angular momentum. 
This approach allows us to define differential operators which act on tensors of any rank. The method is based on irreducible
 tensors of rank $l$ that have $2l+1$ components
and the same transformation properties under rotation as spherical harmonics. Then, from the angular momenta coupling methods 
 \cite{Rotenberg}, irreducible (spherical) tensors of any rank are constructed by using the spherical components of a given set of vector quantities. 
If ${\bf L}={\bf L}_{1}+
{\bf L}_{2}$ is the total angular momentum operator, then the Clebsch-Gordan coefficients are the expansion coefficients in an uncoupled tensor product basis \cite{Cornwell}. Therefore, given any two spherical tensors $T^{(l_{1})}_{m_{1}}$ and $T^{(l_{2})}_{m_{2}}$ of rank $l_{1}$ and 
$l_{2}$, we define a tensor product of these two tensors
\begin{equation}
Q^{(l)}_{m}=\sum_{m_{1}} 
\left(
\begin{array}{ccc}
l_{1} & l_{2} & l \\
m_{1} & m_{2} & m 
\end{array}\right) T^{(l_{1})}_{m_{1}}T^{(l_{2})}_{m_{2}}.
\label{b0}
\end{equation}
This relation, in particular for $l_{1}=1$ and $l_{2}=l$ with $l \rightarrow l-1,l,l+1$, is used to define differential operators in terms of the 
Clebsch-Gordan coefficient using Wigner notation.  

Of special interest in the present work is the ``curl'' operator; it is represented by a 
{\em square} matrix, unlike other operators, like the divergence and the gradient. The curl operator's privileged status  
 is motivated by the elegant properties of square matrices. 
Physical motivation for the group theory approach to the curl operator stems from an 
analogy between electromagnetism and gravity, called gravitoelectromagnetism \cite{Ramos}. This physical content gives rise to a new definition of the curl operator in terms of the angular 
momentum matrices in a Hilbert space. This allows us to construct hermitian 
(or self-adjoint) matrices useful to study the spectra of the curl operator \cite{Yoshida,Picard,Picard1} and its applications to force-free magnetic 
fields \cite{Chandra,Taylor,Montgo,Yoshida1} in plasma physics, astrophysics and fluid dynamics. Moreover, it is possible to extend the standard definition 
of the curl operator in cartesian coordinates to the complex plane.

This paper is organized as follows. In Section 2, we recall \cite{Ramos1}
 the matrix representations of various differential operators (divergence, gradient and curl) which were obtained by using the 
Clebsch-Gordan coefficients, and display some of their properties. In Section 3, we give a new definition and physical
 motivation of the curl operator. In Section 4, we
 study the construction of hermitian matrices for the curl operator, and their extension to the complex plane including some applications. 
Finally, some concluding 
remarks are given in Section 5.

\section{Matrix representations of the differential operators}

Clebsch-Gordan coefficients are  utilized to express the tensor
 product of two irreducible representations of the rotation group, $SO(3)$, as a sum of irreducible 
representations \cite{Cornwell}. We follow Eq. (\ref{b0})  and express
the differential operators: divergence ($\DIV$), gradient ($\GRAD$) and curl ($\CURL$), in the spherical, or canonical, basis \cite{Edmonds} 
through a matrix representation, in terms of the Clebsch-Gordan 
coefficients. The matrix elements which represent the differential operators are given by (see Ref. \cite{Ramos1} and appendix) 
\medskip

{\bf i) Divergence:}
\[
{\DIV}^{(l)}_{m_{1}m_{2}}=-\sqrt{\frac{l(2l+1)}{2l-1}}
\left(
\begin{array}{ccc}
1 & l & l-1 \\
m_{1}-m_{2} & m_{2} & m_{1} 
\end{array}\right)
\partial^{*}_{m_{1}-m_{2}},
\label{a1}
\]    
with $-l+1 \leq m_{1} \leq l-1$,  $-l \leq m_{2} \leq l$, and $\DIV$ is a $(2l-1)\times(2l+1)$ matrix. The operator $\partial^{*}$ is a column vector. 
\medskip

{\bf ii) Gradient:}
\[
{\GRAD}^{(l)}_{m_{1}m_{2}}=
\frac{1}{\sqrt{l+1}}
\left(
\begin{array}{ccc}
1 & l & l+1 \\
m_{1}-m_{2} & m_{2} & m_{1} 
\end{array}\right)
\partial^{*}_{m_{1}-m_{2}},
\label{a2}
\]    
with $-l-1 \leq m_{1} \leq l+1$ and $-l \leq m_{2} \leq l$ and where $\GRAD$ is a $(2l+3)\times(2l+1)$ matrix.  The matrices for the divergence and gradient 
operators, with $l=1$ and $l=2$, are given in Ref. \cite{Ramos1}.
\medskip

{\bf iii) Curl:}  
\begin{equation}
{\CURL}^{(l)}_{m_{1}m_{2}}=i\frac{\sqrt{l(l+1)}}{l}
\left(
\begin{array}{ccc}
1 & l & l \\
m_{1}-m_{2} & m_{2} & m_{1} 
\end{array}\right)
\partial^{*}_{m_{1}-m_{2}},
\label{a3}
\end{equation}     
with $-l \leq m_{1} \leq l$ and $-l \leq m_{2} \leq l$, where $\CURL$ is a $(2l+1)\times(2l+1)$ matrix. 
For instance, for $l=1$, we have 
\begin{equation}
{\CURL}^{(1)}=\frac{1}{i}\left(
\begin{array}{ccc}
\partial_{z} & \frac{\sqrt{2}}{2}(\partial_{x}-i\partial_{y}) & 0 \\
\frac{\sqrt{2}}{2}(\partial_{x}+i\partial_{y}) & 0 & \frac{\sqrt{2}}{2}(\partial_{x}-i\partial_{y}) \\
0 & \frac{\sqrt{2}}{2}(\partial_{x}+i\partial_{y}) & -\partial_{z}
\end{array}\right),
\label{a4}
\end{equation}
and for $l=2$, we get
\begin{eqnarray}
{\CURL}^{(2)}=\frac{1}{2i}\left(
\begin{array}{ccccc}
2\partial_{z} & \partial_{x}-i\partial_{y} & 0 & 0 & 0 \\
\partial_{x}+i\partial_{y} & \partial_{z} & \frac{\sqrt{6}}{2}(\partial_{x}-i\partial_{y}) & 0 & 0 \\
0 & \frac{\sqrt{6}}{2}(\partial_{x}+i\partial_{y}) & 0 & \frac{\sqrt{6}}{2}(\partial_{x}-i\partial_{y}) & 0 \\
0 & 0 & \frac{\sqrt{6}}{2}(\partial_{x}+i\partial_{y}) & -\partial_{z} & \partial_{x}-i\partial_{y} \\
0 & 0 & 0 & \partial_{x}+i\partial_{y} & -2\partial_{z}
\end{array}\right).
\label{b1}
\end{eqnarray}
With these definitions, differential operators can be applied to any spherical tensor $T^{(l)}$ of rank $l$.

Although the matrix structure of  $\DIV$, $\GRAD$ and $\CURL$  is simple and elegant, it is important to note that only the $\CURL$ operator is represented 
by a 
square, $(2l+1)\times (2l+1)$, matrix. This allows us to study these operators further. These three  
operators satisfy the following properties \cite{Ramos1}:
\begin{eqnarray}
{\CURL}^{(1)}{\GRAD}^{(0)}T^{(0)}&=&0,\label{a5} \\ 
{\DIV}^{(1)}{\CURL}^{(1)}T^{(1)}&=&0,  \nonumber \label{a6}\\
{\CURL}^{(1)}{\CURL}^{(1)}T^{(1)}&=&{\GRAD}^{(0)}{\DIV}^{(1)}T^{(1)}-\nabla^{2}T^{(1)},  \nonumber\label{a7}\\
{\CURL}^{(2)}{\GRAD}^{(1)}T^{(1)}&=&\frac{1}{2}{\GRAD}^{(1)}{\CURL}^{(1)}T^{(1)},  \nonumber\label{a8}\\
{\DIV}^{(2)}{\CURL}^{(2)}T^{(2)}&=&\frac{1}{2}{\CURL}^{(1)}{\DIV}^{(2)}T^{(2)}, \nonumber\label{a9}\\ 
{\CURL}^{(l)}{\CURL}^{(l)}T^{(l)}&=&\frac{2l-1}{l}{\GRAD}^{(l-1)}{\DIV}^{(l)}T^{(l)}-\nabla^{2}T^{(l)}. \label{a10}
\end{eqnarray}
Henceforth, we shall concentrate only on the ${\CURL}^{(1)}$ operator, since it is directly connected with the standard curl operator $\nabla \times$  
in cartesian coordinates, ${\mathbb R}^{3}$. Let us
identify some interesting properties of ${\CURL}^{(1)}$. To this end, we split Eq. (\ref{a4}) as follows:
\[
{\CURL}^{(1)}=\left(
\begin{array}{ccc}
0 & -\frac{\sqrt{2}}{2}\partial_{y} & 0 \\
\frac{\sqrt{2}}{2}\partial_{y} & 0 & -\frac{\sqrt{2}}{2}\partial_{y} \\
0 & \frac{\sqrt{2}}{2}\partial_{y} & 0 
\end{array}\right)+i\left(
\begin{array}{ccc}
-\partial_{z} & -\frac{\sqrt{2}}{2}\partial_{x} & 0 \\
-\frac{\sqrt{2}}{2}\partial_{x} & 0 & -\frac{\sqrt{2}}{2}\partial_{x} \\
0 & -\frac{\sqrt{2}}{2}\partial_{x} & \partial_{z}
\end{array}\right).
\label{a11}
\]
The real part of ${\CURL}^{(1)}$ is antisymmetric whereas its imaginary part is symmetric and traceless. Furthermore,
 for the $({\CURL}^{(1)})^{2}$ operator, we have
\begin{eqnarray}\nonumber
({\CURL}^{(1)})^{2}&=&\left(
\begin{array}{ccc}
-\partial^{2}_{z}-\frac{1}{2}(\partial^{2}_{x}+\partial^{2}_{y}) & -\frac{\sqrt{2}}{2}\partial_{x}\partial_{z} & -\frac{1}{2}(\partial^{2}_{x}-
\partial^{2}_{y}) \\
-\frac{\sqrt{2}}{2}\partial_{x}\partial_{z} & -(\partial^{2}_{x}+\partial^{2}_{y}) & \frac{\sqrt{2}}{2}\partial_{x}\partial_{z} \\
-\frac{1}{2}(\partial^{2}_{x}-\partial^{2}_{y}) & \frac{\sqrt{2}}{2}\partial_{x}\partial_{z} & -\partial^{2}_{z}-\frac{1}{2}(\partial^{2}_{x}+
\partial^{2}_{y})
\end{array}\right) \\ 
&+&i\left(
\begin{array}{ccc}
0 & \frac{\sqrt{2}}{2}\partial_{y}\partial_{z} & \partial_{x}\partial_{y} \\
-\frac{\sqrt{2}}{2}\partial_{y}\partial_{z} & 0 & -\frac{\sqrt{2}}{2}\partial_{y}\partial_{z} \\
-\partial_{x}\partial_{y} & \frac{\sqrt{2}}{2}\partial_{y}\partial_{z} & 0 
\end{array}\right).
\label{b2}
\end{eqnarray}
Here real and imaginery parts are symmetric and anti-symmetric respectively. For a general $n$, we find that
\begin{equation}
({\CURL}^{(1)})^{n}=\left\{
\begin{array}{c}
\left(\begin{array}{c} \rm antisymmetric \\ \rm matrix \end{array}\right)+i\left(\begin{array}{c} \rm symmetric\,\ \rm traceless \\ \rm matrix \end{array}\right),\  n \rm{\ odd}, \\
 \\
 \\
\left(\begin{array}{c} \rm symmetric \\ \rm matrix \end{array}\right)+i\left(\begin{array}{c} \rm antisymmetric \\ \rm matrix \end{array}\right), \ 
 n \rm{\ even},
\end{array}\right. 
\label{a12}
\end{equation}

\begin{eqnarray}\nonumber
({\CURL}^{(1)})^{2n}&=&(-1)^{n-1}({\CURL}^{(1)})^{2}\nabla^{2n-2},\nonumber \\
\left({\CURL}^{(1)}\right)^{2n+1}&=&(-1)^{n}({\CURL}^{(1)})\nabla^{2n},\nonumber\label{a14}
\end{eqnarray}
for $n=1, 2, 3, ...$, and
\begin{equation}
e^{{\CURL}^{(1)}}=1+\sum_{n=0}^{\infty}\frac{(-1)^{n}}{(2n+1)!}\left\{{\CURL}^{(1)}+\frac{({\CURL}^{(1)})^{2}}{2n+2}\right\}\nabla^{2n}.
\label{a15}
\end{equation}
These properties can be extended to any curl operator ${\CURL}^{(l)}$ using matrix operations. 

Up to this point, we have discussed the properties of ${\CURL}$ as defined in {\em spherical
 coordinates}.
Now, let us study the above properties within the {\em cartesian coordinates} in ${\mathbb R}^{3}$. According to the theory of harmonic functions 
\cite{Edmonds}, the transformation matrix, $S$, from a spherical, or canonical, basis to a {\it cartesian basis} is given by
\begin{eqnarray}
S=\left(
\begin{array}{ccc}
-\frac{1}{\sqrt{2}} & 0 & \frac{1}{\sqrt{2}} \\
-\frac{i}{\sqrt{2}} & 0 & \frac{i}{\sqrt{2}} \\
0 & 1 & 0
\end{array}\right). \nonumber
\end{eqnarray}
Under this transformation, the ${\CURL}^{(1)}$ operator is equivalent to 
\begin{equation}
\nabla\times=\left(
\begin{array}{ccc}
0 & -\partial_{z} & \partial_{y} \\
\partial_{z} & 0 & -\partial_{x} \\
-\partial_{y} & \partial_{x} & 0
\end{array}\right).
\label{a16}
\end{equation}
This is the matrix form of the standard curl operator in cartesian coordinates. From Eq. (\ref{a16}), we obtain
the following well-known properties:
\begin{eqnarray}\nonumber
\nabla \times \nabla T&=&0, \\ 
\nabla \cdot (\nabla \times {\bf T})&=&0, \\ \nonumber
\nabla \times \nabla \times {\bf T}&=&\nabla(\nabla \cdot {\bf T})-\nabla^{2}{\bf T},
\nonumber
\end{eqnarray}
where $\nabla\equiv(\partial_{x},\partial_{y},\partial_{z})$. These properties are analogous to the properties given by Eqs.
 (\ref{a5}) and (\ref{a7}) for the ${\CURL}^{(1)}$ operator.
Other interesting properties are
\[
[\nabla \times]^{n}=\left\{
\begin{array}{c}
\left(\begin{array}{c} \rm antisymmetric \\ \rm matrix \end{array}\right),\  n {\rm \ odd}, \\
 \\
\left(\begin{array}{c} \rm symmetric \\ \rm matrix \end{array}\right),\ n  \rm{\ even},
\end{array}\right.
\label{a17}
\]

\begin{eqnarray}\nonumber
[\nabla \times]^{2n}&=&(-1)^{n-1}[\nabla \times]^{2}\nabla^{2n-2}, \\
\left[\nabla \times \right]^{2n+1}&=&(-1)^{n}[\nabla \times]\nabla^{2n},
\nonumber
\end{eqnarray}
where $n=1, 2, 3, ...,$ and
\[
e^{\nabla \times}=1+\sum_{n=0}^{\infty}\frac{(-1)^{n}}{(2n+1)!}\left\{\nabla \times+\frac{[\nabla \times]^{2}}{2n+2}\right\}\nabla^{2n}.
\]
In the next section, we discuss physical motivation for this construction of the curl operator. 

\section{Alternative definition of the curl operator}

The electromagnetic fields are determined by Maxwell's equations, which reduce to a wave equation which describes waves 
that are called photons. For each photon
in a momentum state $k$, there are only two degrees of freedom, the helicity (polarization) states, for which the spin is either
oriented along (helicity $+1$) or opposite (helicity $-1$) to the direction of propagation.

The analogy between electromagnetism and gravity can be established by following group-theoretical methods \cite{Ramos}. Inhomogeneous Lorentz group leads
to states that are either covariant or unitary irreducible representations. Using this freedom leads to constrained equations. Then Maxwell-like
equations are derived from these results. In order to clarify the analogy between gravity and electromagnetism, an alternative definition \cite{Ramos}
of the curl operator in Hilbert space was introduced 

\begin{equation}
{\CURL}^{(l)}=\frac{1}{i}\frac{{\bf L}^{(l)} \cdot \nabla}{l},
\label{a21}
\end{equation}
where ${\bf L}^{(l)}$ denotes the angular momentum operator, represented by $(2l+1) \times (2l+1)$ matrix, and ${\bf L}$ is given in the orthonormal 
canonical basis of the Hilbert space. The curl operator in Eq. (\ref{a21}) can be applied to a spherical tensor of rank $l$.  This leads to the following  set of constraints \cite{Ramos}
\begin{eqnarray}
\begin{array}{ll} 
\displaystyle {\CURL}\,\ T^{(l)}_{E}+\frac{1}{c}\frac{\partial T^{(l)}_{B}}{\partial t} =  0, &  \displaystyle {\CURL}\,\ T^{(l)}_{B}-\frac{1}{c}\frac{\partial T^{(l)}_{E}}{\partial t}=0, \\
 & \\
{\DIV}\,\ T^{(l)}_{E} =  0,  & {\DIV}\,\ T^{(l)}_{B}= 0,
\end{array}
\label{b3}
\end{eqnarray}
where $T^{(l)}_{E}$ and $T^{(l)}_{B}$ are rank-$l$ spherical tensors and pseudo-tensors, respectively.   
With $l=1$, Eq. (\ref{b3}) reduces to the free-field Maxwell's 
equations, and, for $l=2$, we obtain the analogous equations, in free space, coupling the gravitoelectric and gravitomagnetic fields.  

It is straightforward to realize that, for $l=1$ and $l=2$, we obtain the same matrices for the ${\CURL}$ operator as in Eqs. (\ref{a4}) and (\ref{b1}),
respectively. Therefore, definitions given by Eqs. (\ref{a3}) and (\ref{a21}) are equivalent and have properties  discussed earlier. 
The definition, Eq. (\ref{a21}), can also be expressed in terms of raising and lowering operators as follows:
\[
{\CURL}^{(l)}=\frac{1}{il}\left[L_{z}^{(l)}\partial_{z}+\frac{1}{2}(L_{+}^{(l)}\partial_{-}+L_{-}^{(l)}\partial_{+})\right],
\label{a22}
\]
where $L_{\pm}=L_{x}\pm iL_{y}$ and $\partial_{\pm}=\partial_{x}\pm \partial_{y}$. This shows the connection between the curl operator and the 
irreducible representations of the proper rotation group $SO(3)$. 

\section{Complex curl operator}

The ${\CURL}^{(l)}$ operator, defined in Eq. (\ref{a21}), is antihermitian. The hermitian curl operator, 
${\CURL}^{(l)}_{H}$, is written as ${\CURL}^{(l)}_{H}\equiv i\ {\CURL}^{(l)}$. In particular, for $l=1$, we have 
\[
{\CURL}^{(1)}_{H}=\left(
\begin{array}{ccc}
\partial_{z} & \frac{\sqrt{2}}{2}(\partial_{x}-i\partial_{y}) & 0 \\
\frac{\sqrt{2}}{2}(\partial_{x}+i\partial_{y}) & 0 & \frac{\sqrt{2}}{2}(\partial_{x}-i\partial_{y}) \\
0 & \frac{\sqrt{2}}{2}(\partial_{x}+i\partial_{y}) & -\partial_{z}
\end{array}\right).
\label{a23}
\]
Some properties of this hermitian curl-operator are
\begin{eqnarray}\nonumber
{\CURL}^{(1)}_{H}{\GRAD}^{(0)}T^{(0)}&=&0, \nonumber\\
{\DIV}^{(1)}{\CURL}^{(1)}_{H}T^{(1)}&=&0, \nonumber\\ \label{a25}
{\CURL}^{(1)}_{H}{\CURL}^{(1)}_{H}T^{(1)}&=&\nabla^{2}T^{(1)}-{\GRAD}^{(0)}{\DIV}^{(1)}T^{(1)},\nonumber \\ \label{a26}
{\CURL}^{(2)}_{H}{\GRAD}^{(1)}T^{(1)}&=&\frac{1}{2}{\GRAD}^{(1)}{\CURL}^{(1)}_{H}T^{(1)}, \nonumber\\ \label{a27}
{\DIV}^{(2)}{\CURL}^{(2)}_{H}T^{(2)}&=&\frac{1}{2}{\CURL}^{(1)}_{H}{\DIV}^{(2)}T^{(2)}, \nonumber\\ \label{a28}
{\CURL}^{(l)}_{H}{\CURL}^{(l)}_{H}T^{(l)}&=&\nabla^{2}T^{(l)}-\frac{2l-1}{l}{\GRAD}^{(l-1)}{\DIV}^{(l)}T^{(l)}. \nonumber
\label{a29}
\end{eqnarray}

Let us define a new curl operator, ${\sCURL}^{(1)}$, in terms of ${\CURL}^{(1)}_{H}$ and ${\CURL}^{(1)}$ as 
\[
\begin{array}{rcl}
{\sCURL}^{(1)}&=& {\CURL}^{(1)}_{H}+ {\CURL}^{(1)}\\
&=&(1-i)
\left(
\begin{array}{ccc}
\partial_{z} & \frac{\sqrt{2}}{2}(\partial_{x}-i\partial_{y}) & 0 \\
\frac{\sqrt{2}}{2}(\partial_{x}+i\partial_{y}) & 0 & \frac{\sqrt{2}}{2}(\partial_{x}-i\partial_{y}) \\
0 & \frac{\sqrt{2}}{2}(\partial_{x}+i\partial_{y}) & -\partial_{z}
\end{array}\right).\end{array}
\label{a30}
\]
The properties of the $\sCURL$ operator are
\begin{eqnarray}\nonumber
{\sCURL}^{(1)}{\GRAD}^{(0)}T^{(0)}&=&0,\nonumber \\
{\DIV}^{(1)}{\sCURL}^{(1)}T^{(1)}&=&0,\nonumber \\ \label{a32}
{\sCURL}^{(1)}{\sCURL}^{(1)}T^{(1)}&=&2i({\GRAD}^{(0)}{\DIV}^{(1)}T^{(1)}-\nabla^{2}T^{(1)}), \nonumber\\ \label{a33}
{\sCURL}^{(2)}{\GRAD}^{(1)}T^{(1)}&=&\frac{1}{2}{\GRAD}^{(1)}{\sCURL}^{(1)}T^{(1)},\nonumber \\ \label{a34}
{\DIV}^{(2)}{\sCURL}^{(2)}T^{(2)}&=&\frac{1}{2}{\sCURL}^{(1)}{\DIV}^{(2)}T^{(2)}, \nonumber\\ \label{a35}
{\sCURL}^{(l)}{\sCURL}^{(l)}T^{(l)}&=&2i\left(\frac{2l-1}{l}{\GRAD}^{(l-1)}{\DIV}^{(l)}T^{(l)}-\nabla^{2}T^{(l)}\right). \nonumber
\label{a36}
\end{eqnarray}

In cartesian coordinates, the $\sCURL$ operator becomes 
\[
\nabla_{c} \times = \left(
\begin{array}{ccc}
0 & -\partial_{z} & \partial_{y} \\
\partial_{z} & 0 & -\partial_{x} \\
-\partial_{y} & \partial_{x} & 0
\end{array}\right)+i\left(
\begin{array}{ccc}
0 & -\partial_{z} & \partial_{y} \\
\partial_{z} & 0 & -\partial_{x} \\
-\partial_{y} & \partial_{x} & 0
\end{array}\right). 
\label{a37}
\] 
If we write the last equation in terms of the standard curl operator, $\nabla \times$, we find
\begin{equation}
\nabla_{c} \times = \nabla \times +\; i\;\nabla \times.
\label{a38}
\end{equation}
This defines the extension of the curl operator to the complex plane. Note that
\[
\nabla_{H} \times = i\nabla \times
\label{a39}
\]
is hermitian. 
Some properties of the operators $\nabla_{c} \times $ and $\nabla_{H} \times$ are 
\begin{eqnarray}\nonumber\label{a40}
\nabla_{c} \times \nabla\,\ T&=&0,\nonumber \\
\nabla \cdot (\nabla_{c} \times {\bf T})&=&0,\nonumber\\ \label{a41}
\nabla_{c} \times \nabla_{c} \times {\bf T}&=&2i[\nabla(\nabla \cdot {\bf T})-\nabla^{2}{\bf T}],\nonumber\\ \label{a42}
\nabla_{H} \times \nabla\,\ T&=&0,\nonumber \\ \label{a43}
\nabla \cdot (\nabla_{H} \times {\bf T})&=&0,\nonumber\\ \label{a44}
\nabla_{H} \times \nabla_{H} \times {\bf T}&=&\nabla^{2}{\bf T}-\nabla(\nabla \cdot {\bf T}).\nonumber
\label{a45}
\end{eqnarray}

Next, we provide a simple example that show how to use the complex curl operator and more sophisticated applications of this operator. 

\newpage

\vspace{0.5cm}
{\bf Example 1}

\vspace{0.3cm}
Let ${\bf F}(x,y,z)$ be a complex vector field
\begin{equation}
{\bf F}(x,y,z)={\bf u}(x,y,z)+i{\bf v}(x,y,z),
\label{a461}
\end{equation}
then, by using Eq.(\ref{a38}), the complex curl of ${\bf F}(x,y,z)$ is
\begin{equation}
\nabla_{c} \times {\bf F}=\nabla \times ({\bf u}-{\bf v})+i\nabla \times ({\bf u}+{\bf v}).
\label{a462}
\end{equation}
Let us take a specific form for ${\bf F}$
\begin{eqnarray}
{\bf u}(x,y,z)=y{\bf i}-x{\bf j}, \,\,\,\,\,\,\,\,\ {\bf v}(x,y,z)=-x^{2}{\bf j},
\nonumber
\end{eqnarray} 
where ${\bf u}$ and ${\bf v}$ represent, separately, fields rotating on the plane XY. Substituting ${\bf u}$ and {\bf v} in Eq.(\ref{a462}) it yields
\begin{equation}
\nabla_{c} \times {\bf F}=2(x-1){\bf k}-2i(x+1){\bf k}.
\label{463}
\end{equation}
Notice that $\nabla_{c} \times {\bf F}$ is along the $Z$ direction as expected.

\vspace{0.5cm}
{\bf Example 2}

\vspace{0.3cm}
Let us express the source-free Maxwell's equations in terms of the complex curl operator, $\nabla_{c} \times$, as
\begin{eqnarray}
\begin{array}{ll} 
 \displaystyle\nabla_{c} \times {\bf E}+\frac{1}{c}\frac{\partial {\bf B}}{\partial t}+\frac{i}{c}\frac{\partial {\bf B}}{\partial t} =  0,  &  
 \displaystyle\nabla_{c} \times {\bf B}-\frac{1}{c}\frac{\partial {\bf E}}{\partial t}-\frac{i}{c}\frac{\partial {\bf E}}{\partial t}=0, \\
 &\nonumber \\
\nabla \cdot {\bf E} =  0,  & \nabla \cdot {\bf B}= 0.
\end{array}
\label{a46}
\end{eqnarray}
For gravitoelectromagnetism, following these arguments and those presented in Section 2 for the ${\CURL}^{(2)}$ operator, 
we write the source-free Maxwell-like 
equations of gravitoelectromagnetism \cite{Ramos}  as

\begin{eqnarray}
\begin{array}{ccccccc} 
\displaystyle {\curl}_{c}\, {\cal E}+\frac{1}{c}\frac{\partial {\cal B}}{\partial t}+\frac{i}{c}\frac{\partial {\cal B}}{\partial t}& = & 0, & &  
\displaystyle {\curl}_{c}\, {\cal B}-\frac{1}{c}\frac{\partial {\cal E}}{\partial t}-\frac{i}{c}\frac{\partial {\cal E}}{\partial t}&=&0, \\ \nonumber
 & \\
{\tdiv}\, {\cal E}& = & 0, & & {\tdiv}\, {\cal B}&=& 0,
\end{array}
\label{a47}
\end{eqnarray}
where ${\cal E}$ and ${\cal B}$ are symmetric and traceless second rank field tensors. The complex curl 
operator, ${\curl}_{c}$, is given by ${\curl}_{c}={\curl}+i\; {\curl}$, where the $\curl$ operator is defined as
\[
{\curl}\, T =\left(
\begin{array}{ccc}
\partial^2T^{13}-\partial^3T^{12} & a & b \\
 & & \\
a & \partial^3T^{12}-\partial^1T^{23} & c \\
 & & \\
b & c & \partial^1T^{23}-\partial^2T^{13}\end{array}\right),\nonumber
\label{a48}
\]
where
\begin{eqnarray}
a&=&\frac 12\left[(\partial^2T^{32}-\partial^3T^{22})+(\partial^3T^{11}-\partial^1T^{31})\right], \nonumber \\ 
b&=&\frac 12\left[\partial^2T^{33}-\partial^3T^{23})+(\partial^1T^{21}-\partial^2T^{11})\right], \nonumber \\
c&=&\frac 12\left[\partial^3T^{13}-\partial^1T^{33})+(\partial^1T^{22}-\partial^2T^{12})\right].
\nonumber
\end{eqnarray}
Furthermore,$ (T^{ij}), i,j=1,2,3$ is a symmetric and traceless tensor of rank 2. 

\vspace{0.5cm}
{\bf Example 3}

\vspace{0.3cm}
The complex curl operator, $\nabla_{c}\times$, suggests a way to study the Helmholtz decomposition in a complex space. Since the operator $\nabla$, in 
${\bf r}$-space, maps into $i{\bf k}$, in ${\bf k}$-space, then Eq. (\ref{a38}) is written as
\[
\nabla_{c}\times = i{\bf k}_{c}\times,\nonumber
\]
where ${\bf k}_{c}=(1+i){\bf k}$. Thus, $\nabla_{c}\times$ operator, in ${\bf r}$-space, maps into $i{\bf k}_{c}\times$, in ${\bf k}_{c}$-space. 

Let ${\bf F}({\bf r})$ be a vector field in ${\bf r}$-space. We define the Fourier transform of such a vector field by
\begin{eqnarray}\label{a51}
{\bf \tilde{\bf F}}({\bf k}_{c})&=&\int \frac{d^{3}{\bf r}}{(2\pi)^{3/2}}e^{-i{\bf k}_{c}\cdot {\bf r}}{\bf F}({\bf r}), \\
{\bf F}({\bf r})&=&\int \frac{d^{3}{\bf k}_{c}}{(2\pi)^{3/2}}e^{i{\bf k}_{c}\cdot {\bf r}}\tilde{{\bf F}}({\bf k}_{c}),
\label{a52}
\end{eqnarray}
where we use tilde to denote quantities in ${\bf k}_{c}$-space. In addition Eq. (\ref{a51}) becomes
\[
{\bf \tilde{F}}({\bf k}_{c})={\bf \tilde{\cal F}}({\bf k}),
\label{a53}
\]
where
\[
{\bf \tilde{\cal F}}({\bf k})=\int \frac{d^{3}{\bf r}}{(2\pi)^{3/2}}e^{-i{\bf k}\cdot {\bf r}}{\bf \cal F}({\bf r}),
\label{a54}
\]
with ${\bf \cal F}({\bf r})=e^{{\bf k}\cdot {\bf r}}{\bf F}({\bf r})$. On the other hand, Eq. (\ref{a52}) is expressed as
\begin{equation}
{\bf F}({\bf r})=(1+i){\bf G}({\bf r}),
\label{a55}
\end{equation}
where ${\bf G}({\bf r})$ is a real vector field,  given by
\begin{eqnarray}\label{a56}
{\bf G}({\bf r})&=&\int \frac{d^{3}{\bf k}}{(2\pi)^{3/2}}e^{i{\bf k}\cdot {\bf r}}{\bf \tilde{G}}({\bf k}),\\
{\bf \tilde{G}}({\bf k})&=&\int \frac{d^{3}{\bf r}}{(2\pi)^{3/2}}e^{-i{\bf k}\cdot {\bf r}}{\bf G}({\bf r}),
\label{a57}
\end{eqnarray}
with ${\bf \tilde{G}}({\bf k})=e^{-{\bf k}\cdot {\bf r}}{\bf \tilde{\cal F}}({\bf k})$. Note that Eqs. (\ref{a56}) and (\ref{a57}) are the Fourier transforms
of the real vector field ${\bf G}({\bf r})$ in ${\bf k}$-space.

From Eq. (\ref{a55}), we conclude that ${\bf F}({\bf r})$ is a complex vector field with its Fourier transform given as ${\bf \tilde{\bf F}}({\bf k}_{c})$.  
Both fields are written in terms of the real vector field, ${\bf G}({\bf r})$, and its Fourier transformed field, ${\bf \tilde{G}}({\bf k})$,  as
\begin{eqnarray}\label{a58}
{\bf F}({\bf r})&=&(1+i){\bf G}({\bf r}),\nonumber\\
{\bf \tilde{\bf F}}({\bf k}_{c})&=&e^{{\bf k}\cdot {\bf r}}{\bf \tilde{G}}({\bf k}).\nonumber
\label{59}
\end{eqnarray}
The Helmholtz decomposition of a real vector field ${\bf G}({\bf r})$ into its longitudinal ${\bf G}_{\parallel}({\bf r})$ and transverse 
${\bf G}_{\perp}({\bf r})$ components is well known \cite{Morse}. Using this procedure, we decompose the complex vector field ${\bf F}({\bf r})$ into its 
longitudinal and transverse components as 
\begin{equation}
{\bf \tilde{\bf F}}({\bf k}_{c})={\bf \tilde{\bf F}}_{\perp}({\bf k}_{c})+{\bf \tilde{\bf F}}_{\parallel}({\bf k}_{c})=e^{{\bf k}\cdot {\bf r}}
{\bf \tilde{G}}_{\perp}({\bf k})+e^{{\bf k}\cdot {\bf r}}{\bf \tilde{G}}_{\parallel}({\bf k}),
\label{a60}
\end{equation}
with
\begin{eqnarray}
{\bf \tilde{\bf F}}_{\parallel}({\bf k}_{c})&=&-i\frac{{\bf k}_{c}}{k^{2}_{c}}({\bf k}_{c}\cdot {\bf \tilde{\bf F}}({\bf k}_{c})),\nonumber \\ 
{\bf k}\cdot {\bf \tilde{\bf F}}_{\perp}({\bf k}_{c})&=&0,\label{a63}\\ 
{\bf k}_{c}\cdot {\bf \tilde{\bf F}}_{\parallel}({\bf k}_{c})&=&0.\nonumber
\end{eqnarray}
Upon transforming Eqs. (\ref{a60})-(\ref{a63}) back to ${\bf r}$-space, we obtain the longitudinal and transverse components of ${\bf F}({\bf r})$ 
as 
\[
{\bf F}({\bf r})={\bf F}_{\perp}({\bf r})+{\bf F}_{\parallel}({\bf r})=(1+i){\bf G}_{\perp}({\bf r})+(1+i){\bf G}_{\perp}({\bf r}),
\label{a64}
\]
with
\begin{eqnarray}\label{a65}
\nabla\cdot {\bf F}_{\perp}({\bf r})=0,\nonumber \\
\nabla_{c}\times {\bf F}_{\parallel}({\bf r})=0.\nonumber
\label{a66}
\end{eqnarray}
Thus, a complex vector field ${\bf F}({\bf r})$ is decomposed into a divergence-free part, ${\bf F}_{\perp}({\bf r})$, and a curl-free part, 
 ${\bf F}_{\parallel}({\bf r})$. 

\section{Concluding remarks}

In this paper, a group-theoretical matrix approach of the curl operator is discussed. It is 
 physically motivated by an analogy between the field 
equations describing gravity and electromagnetism. The properties for the curl operator, obtained within the matrix formulation, are elegant and simple.
These are reminiscent of the traditional curl operator. They reflect the symmetry of the matrix procedure. The alternative definition of the curl operator 
provides a way to construct hermitian 
representations which is applied in the study of the spectra of the curl operator. 

This approach suggests, in a natural way, that the definition of the curl operator can be extended to the complex
 plane. This operator may be also applied to complex 
vector fields, in general, and the electromagnetic field, in particular. 

\vspace{0.5cm}

\section*{Acknowlegdement} The authors are grateful to the Natural Sciences and Engineering Research Council  (NSERC, Canada) for financial support.

\section*{Appendix}

Hereafter, we recall some important relations from Ref. \cite{Ramos1}.  The general expression for the {\it {divergence}} is given by Eq. (3) of Ref. \cite{Ramos1}:
\begin{equation}
({\mathrm{div}}\ T)_{\underbrace{bcd\cdots}_{l-1}}=-\sqrt{\frac{l(2l+1)}{2l-1}}\sum_{m_1=1,0,-1}
\left(
\begin{array}{ccc}
1 & l & l-1 \\
m_{1} & m_{2} & m 
\end{array}\right)
\partial_{m_1}T_{m_2}^{(l)}.
\label{appDiv}
\end{equation}   
The general expression for the {\it {gradient}} is in Eq. (9) of the same paper,
\begin{equation}
({\mathrm{grad}}\ T)_{\underbrace{abcd\cdots}_{l+1}}=
\frac{1}{\sqrt{l+1}}\sum_{m_1=1,0,-1}
\left(
\begin{array}{ccc}
1 & l & l+1 \\
m_{1} & m_{2} & m 
\end{array}\right)
\partial_{m_1}T_{m_2}^{(l)},
\label{appGrad}
\end{equation}  
while the relation for the {\it {curl}} is given by Eq. (6) of the same paper,
\begin{equation}
({\mathrm{curl}}\ T)_{\underbrace{abc\cdots}_{l}}=
-i\frac{\sqrt{l(l+1)}}{l}\sum_{m_1=1,0,-1}
\left(
\begin{array}{ccc}
1 & l & l \\
m_{1} & m_{2} & m 
\end{array}\right)
\partial_{m_1}T_{m_2}^{(l)},
\label{appCurl}
\end{equation}  

With $l=1$, the gradient operator has the form given by Eq. (14) of Ref. \cite{Ramos1}:
\[
GRAD^{(1)}=\left (\begin{array}{ccc}
-\frac12(\partial_x-i\partial_y) & 0 & 0\\
\frac 12\partial_z & -\frac1{2\sqrt{2}}(\partial_x-i\partial_y) & 0\\
\frac1{2\sqrt{6}}(\partial_x+i\partial_y) & \frac 1{\sqrt{3}}\partial_z & -\frac1{2\sqrt{6}}(\partial_x-i\partial_y)\\
0 & \frac1{2\sqrt{2}}(\partial_x+i\partial_y) & \frac 12\partial_z\\
0 & 0 & \frac12(\partial_x+i\partial_y).
\end{array}\right)
\]
The divergence operator, with $l=2$, is given in Eq. (11) of Ref. \cite{Ramos1},
\[
DIV^{(2)}=\left (\begin{array}{ccccc}
-(\partial_x+i\partial_y) & \partial_z & \frac 1{\sqrt{6}}(\partial_x-i\partial_y) & 0 & 0\\
0 & -\frac 1{\sqrt{2}}(\partial_x-i\partial_y)  & \frac 2{\sqrt{3}}\partial_z & \frac 1{\sqrt{2}}(\partial_x-i\partial_y) & 0\\
0 & 0 & -\frac 1{\sqrt{6}}(\partial_x-i\partial_y) & \partial_z & (\partial_x-i\partial_y).
\end{array}\right)
\]

\end{document}